\documentclass[dvips,apjl]{emulateapj}

\usepackage{apjfonts}

\begin{document}

\title{Theoretical interpretation of luminosity and spectral properties of GRB 031203.}

\author{Maria Grazia Bernardini\altaffilmark{1,2}, Carlo Luciano Bianco\altaffilmark{1,2}, Pascal Chardonnet\altaffilmark{1,3}, Federico Fraschetti\altaffilmark{1,4}, Remo Ruffini\altaffilmark{1,2}, She-Sheng Xue\altaffilmark{1,2}}

\altaffiltext{1}{ICRA - International Center for Relativistic Astrophysics. E-mails: ruffini@icra.it, xue@icra.it}
\altaffiltext{2}{Dipartimento di Fisica, Universit\`a di Roma ``La Sapienza", Piazzale Aldo Moro 5, I-00185 Roma, Italy. E-mails: maria.bernardini@icra.it, bianco@icra.it}
\altaffiltext{3}{Universit\'e de Savoie, LAPTH - LAPP, BP 110, F-74941 Annecy-le-Vieux Cedex, France. E-mail: chardon@lapp.in2p3.fr}
\altaffiltext{4}{Universit\`a di Trento, Via Sommarive 14, I-38050 Povo (Trento), Italy. E-mail: fraschetti@icra.it}

\begin{abstract}
The X and $\gamma$-ray observations of the source GRB 031203 by INTEGRAL are interpreted within our theoretical model. In addition to a complete space-time parametrization of the GRB, we specifically assume that the afterglow emission originates from a thermal spectrum in the co-moving frame of the expanding baryonic matter shell. By determining the two free parameters of the model and estimating the density and filamentary structure of the ISM, we reproduce the observed luminosity in the $20-200$ keV energy band. As in previous sources, the prompt radiation is shown to coincide with the peak of the afterglow and the luminosity substructure are shown to originate in the filamentary structure of the ISM. We predict a clear hard-to-soft behavior in the instantaneous spectra. The time-integrated spectrum over $20$ seconds observed by INTEGRAL is well fitted. Despite this source has been considered ``unusual'', it appears to us a normal low energetic GRB.
\end{abstract}


\section{Introduction}

GRB 031203 was observed by IBIS, on board of the INTEGRAL satellite \citep{mg}, as well as by XMM \citep{XMM} and Chandra \citep{sod} in the $2-10$ keV band, and by VLT \citep{sod} in the radio band. It appears as a typical long burst \citep{saz}, with a simple profile and a duration of $\approx 40$ s. The burst fluence in the $20-200$ keV band is $(2.0\pm 0.4)\times 10^{-6}$ erg/cm$^2$ \citep{saz}, and the measured redshift is $z=0.106$ \citep{proch}. We analyze in the following the gamma-ray signal received by INTEGRAL. The observations in other wavelengths, in analogy with the case of GRB 980425 \citep{pian,cospar02}, could be related to the supernova event, as also suggested by \citet{sod}, and they will be examined elsewhere.

The INTEGRAL observations find a direct explanation in our theoretical model \citep[see][and references therein]{rlet1,rlet2,rubr,rubr2,EQTS_ApJL2,PowerLaws}. We determine the values of the two free parameters which characterize our model: the total energy stored in the Dyadosphere $E_{dya}$ and the mass of the baryons left by the collapse $M_Bc^2 \equiv B E_{dya}$. We follow the expansion of the pulse, composed by the electron-positron plasma initially created by the vacuum polarization process in the Dyadosphere. The plasma self-propels itself outward and engulfs the baryonic remnant left over by the collapse of the progenitor star. As such pulse reaches transparency, the Proper Gamma-Ray Burst (P-GRB) is emitted \citep{rswx99,rswx00,rlet2}. The remaining accelerated baryons, interacting with the interstellar medium (ISM), produce the afterglow emission. The ISM is described by the two additional parameters of the theory: the average particle number density $<n_{ISM}>$ and the ratio $<\mathcal{R}>$ between the effective emitting area and the total area of the pulse \citep{spectr1}, which take into account the ISM filamentary structure \citep{fil}.

We reproduce correctly in several GRBs and in this specific case (see e.g. Fig. \ref{fig1}) the observed time variability of the prompt emission \citep[see e.g.][and references therein]{r02,rubr,rubr2}. The radiation produced by the interaction of the accelerated baryons with the ISM agrees with observations both for intensity and time structure.

The progress in reproducing the X and $\gamma-$ray emission as originating from a thermal spectrum in the co-moving frame of the burst \citep{spectr1} leads to the characterization of the instantaneous spectral properties which are shown to drift from hard to soft during the evolution of the system. The convolution of these instantaneous spectra over the observational time scale is in very good agreement with the observed power-law spectral shape.

As shown in previous cases \citep[see][]{rubr,beam}, also for GRB 031203, using the correct equations of motion, there is no need to introduce a collimated emission to fit the afterglow observations \citep[see also][who find this same conclusion starting from different considerations]{sod}.

\section{The initial conditions}

The best fit of the observational data leads to a total energy of the Dyadosphere $E_{dya}=1.85\times10^{50}$ erg. Assuming a black hole mass $M=10M_{\odot}$, we then have a black hole charge to mass ratio $\xi=6.8\times 10^{-3}$; the plasma is created between the radii $r_1=2.95\times10^6$ cm and $r_2=2.81\times10^7$ cm with an initial temperature of $1.52$ MeV and a total number of pairs $N_{e^+e^-}=2.98\times10^{55}$. The amount of baryonic matter in the remnant is $B = 7.4\times10^{-3}$.

After the transparency point and the P-GRB emission, the initial Lorentz gamma factor of the accelerated baryons is $\gamma=132.8$ at an arrival time at the detector $t^d_a=8.14\times 10^{-3}$ s and a distance from the Black Hole $r=6.02\times 10^{12}$ cm. This corresponds to an apparent superluminal velocity along the line of sight of $2.5 \times 10^4c$. The ISM parameters are: $<n_{ism}>=0.3$ particle/$cm^3$ and $<\mathcal{R}>=7.81\times 10^{-9}$.

\section{The GRB luminosity in fixed energy bands}\label{par3}

The aim of our model is to derive from first principles both the luminosity in selected energy bands and the time resolved/integrated spectra. The luminosity in selected energy bands is evaluated integrating over the equitemporal surfaces \citep[EQTSs, see][]{EQTS_ApJL,EQTS_ApJL2} the energy density released in the interaction of the accelerated baryons with the ISM measured in the co-moving frame, duly boosted in the observer frame. The radiation viewed in the co-moving frame of the accelerated baryonic matter is assumed to have a thermal spectrum and to be produced by the interaction of the ISM with the front of the expanding baryonic shell.

In order to evaluate the contributions in the band $[\nu_1,\nu_2]$ we have to multiply the bolometric luminosity with an ``effective weight'' $W(\nu_1,\nu_2,T_{arr})$, where $T_{arr}$ is the observed temperature. $W(\nu_1,\nu_2,T_{arr})$ is given by the ratio of the integral over the given energy band of a Planckian distribution at temperature $T_{arr}$ to the total integral $aT_{arr}^4$ \citep{spectr1}. The resulting expression for the emitted luminosity is
\begin{equation}
\frac{dE_\gamma^{\left[\nu_1,\nu_2\right]}}{dt_a^d d \Omega } = \int_{EQTS} \frac{\Delta \varepsilon}{4 \pi} \; v \; \cos \vartheta \; \Lambda^{-4} \; \frac{dt}{dt_a^d} W\left(\nu_1,\nu_2,T_{arr}\right) d \Sigma\, ,
\label{fluxarrnu}
\end{equation}
where $\Delta \varepsilon=\Delta E_{int}/V$ is the energy density released in the interaction of the accelerated baryons with the ISM measured in the co-moving frame, $\Lambda=\gamma(1-(v/c)\cos\vartheta)$ is the Doppler factor, $d\Sigma$ is the surface element at detector arrival time $t_a^d$ on which the integration is performed \citep[details in][]{spectr1}

\section{The GRB 031203 ``prompt emission''}

\begin{figure}
\includegraphics[width=\hsize,clip]{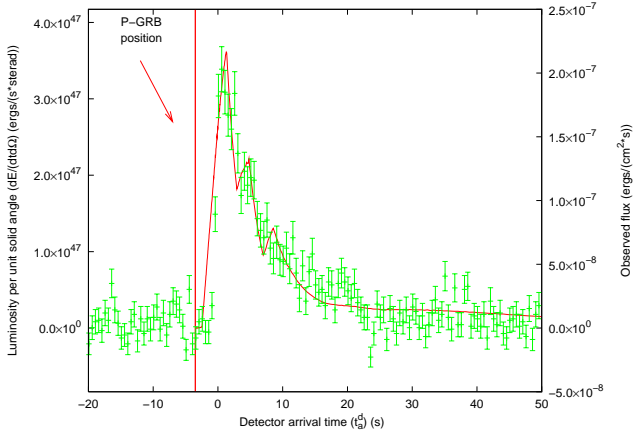}
\caption{Theoretically simulated light curve of the GRB 031203 prompt emission in the $20-200$ keV energy band (solid red line) is compared with the observed data (green points) from \citet{saz}. The vertical bold red line indicates the time position of P-GRB.}
\label{fig1}
\end{figure}

In order to compare our theoretical prediction with the observations, it is important to notice that there is a shift between the initial time of the GRB event and the moment in which the satellite instrument has been triggered. In fact, in our model the GRB emission starts at the transparency point when the P-GRB is emitted. If the P-GRB is under the threshold of the instrument, the trigger starts a few seconds later with respect to the real beginning of the event. Therefore it is crucial, in the theoretical analysis, to estimate and take into due account this time delay. In the present case it results in $\Delta t^d_a=3.5$ s (see the bold solid line in Fig. \ref{fig1}). In what follows, the detector arrival time is referred to the onset of the instrument.

The structure of the prompt emission of GRB 031203, which is a single peak with a slow decay, is reproduced assuming an ISM which has not a constant density but presents several density spikes with $<n_{ISM}>=0.16$ particle/cm$^3$. Such density spikes corresponding to the main peak are modeled as three spherical shells with width $\Delta$ and density contrast $\Delta n/n$: we adopted for the first peak $\Delta=3.0\times10^{15}$ cm and $\Delta n/n=8$, for the second peak $\Delta=1.0\times10^{15}$ cm and $\Delta n/n=1.5$ and for the third one $\Delta=7.0\times10^{14}$ cm and $\Delta n/n=1$. To describe the details of the ISM filamentary structure we would require an intensity vs. time information with an arbitrarily high resolving power. With the finite resolution of the INTEGRAL instrument, we can only describe the average density distribution compatible with the given accuracy. Only structures at scales of $10^{15}$ cm can be identified. Smaller structures would need a stronger signal and/or a smaller time resolution of the detector. The three clouds here considered are necessary and sufficient to reproduce the observed light curve: a smaller number would not fit the data, while a larger number is unnecessary and would be indeterminable.

The result (see Fig. \ref{fig1}) shows a good agreement with the light curve reported by \citet{saz}, and it provides a further evidence for the possibility of reproducing light curves with a complex time variability through ISM inhomogeneities \citep{r02,rubr,rubr2} \citep[see also the analysis of the prompt emission of GRB 991216 in][]{r02}.

\section{The GRB 031203 instantaneous spectrum}\label{inst}

\begin{figure}
\includegraphics[width=\hsize,clip]{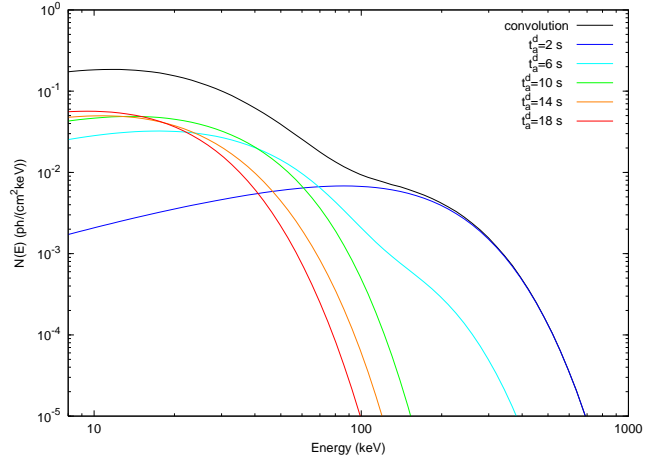}
\caption{Five different theoretically predicted instantaneous photon number spectrum $N(E)$ for $t_a^d=2$, $6$, $10$, $14$, $18$ s are here represented (colored curves) together with their own temporal convolution (black bold curve). The shapes of the instantaneous spectra are not blackbodies due to the spatial convolution over the EQTS (see text).}
\label{fig2}
\end{figure}

As outlined in section \ref{par3}, in addition to the the luminosity in fixed energy bands we can derive also the instantaneous photon number spectrum $N(E)$. In Fig. \ref{fig2} are shown samples of time-resolved spectra for five different values of the arrival time which cover the whole duration of the event.

It is manifest from this picture that, although the spectrum in the co-moving frame of the expanding pulse is thermal, the shape of the final spectrum in the laboratory frame is clearly non thermal. In fact, as explained in \citet{spectr1}, each single instantaneous spectrum is the result of an integration of hundreds of thermal spectra over the corresponding EQTS. This calculation produces a non thermal instantaneous spectrum in the observer frame (see Fig. \ref{fig2}).

\begin{figure}
\includegraphics[width=\hsize,clip]{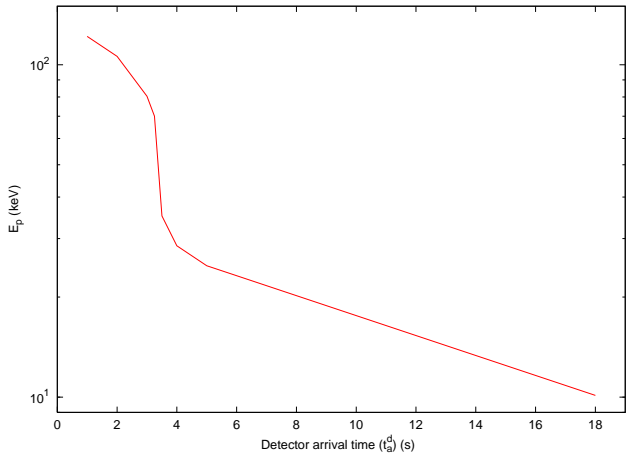}
\caption{The energy of the peak of the instantaneous photon number spectrum $N(E)$ is here represented as a function of the arrival time during the ``prompt emission'' phase. The clear hard to soft behavior is shown.}
\label{fig3}
\end{figure}

Another distinguishing feature of the GRBs spectra which is also present in these instantaneous spectra, as shown in Fig. \ref{fig2}, is the hard to soft transition during the evolution of the event \citep{cri97,p99,fa00,gcg02}. In fact the peak of the energy distributions $E_p$ drift monotonically to softer frequencies with time (see Fig. \ref{fig3}). This feature explains the change in the power-law low energy spectral index $\alpha$ \citep{b93} which at the beginning of the prompt emission of the burst ($t_a^d=2$ s) is $\alpha=0.75$, and progressively decreases for later times (see Fig. \ref{fig2}). In this way the link between $E_p$ and $\alpha$ identified by \citet{cri97} is explicitly shown. This theoretically predicted evolution of the spectral index during the event unfortunately cannot be detected in this particular burst by INTEGRAL because of the not sufficient quality of the data \citep[poor photon statistics, see][]{saz}.

\section{The GRB 031203 time-integrated spectrum and the comparison with the observed data}

The time-integrated observed GRB spectra show a clear power-law behavior. Within a different framework Shakura, Sunyaev and Zel'dovich \citep[see e.g.][and references therein]{poz83} argued that it is possible to obtain such power-law spectra from a convolution of many non power-law instantaneous spectra evolving in time. This result was recalled and applied to GRBs by \citet{bk99} assuming for the instantaneous spectra a thermal shape with a temperature changing with time. They showed that the integration of such energy distributions over the observation time gives a typical power-law shape possibly consistent with GRB spectra.

Our specific quantitative model is more complicated than the one considered by \citet{bk99}: as pointed out in section \ref{inst}, the instantaneous spectrum here is not a black body. Each instantaneous spectrum is obtained by an integration over the corresponding EQTS \citep{EQTS_ApJL,EQTS_ApJL2}: it is itself a convolution, weighted by appropriate Lorentz and Doppler factors, of $\sim 10^6$ thermal spectra with variable temperature. Therefore, the time-integrated spectra are not plain convolutions of thermal spectra: they are convolutions of convolutions of thermal spectra (see Fig. \ref{fig2}).

\begin{figure}
\includegraphics[width=\hsize,clip]{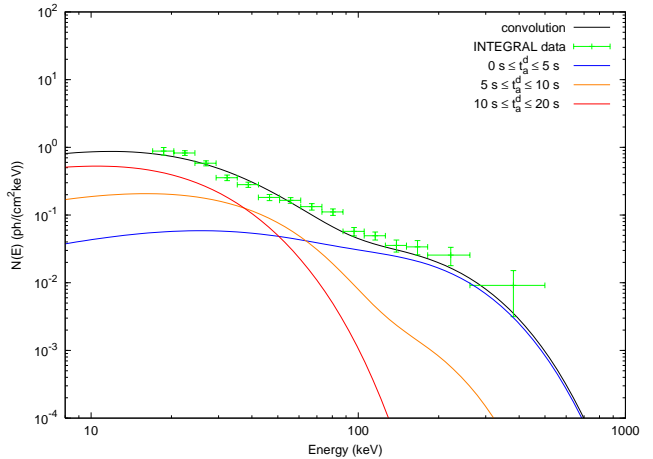}
\caption{Three theoretically predicted time-integrated photon number spectra $N(E)$ are here represented for $0 \le t_a^d \le 5$ s, $5 \le t_a^d \le 10$ s and $10 \le t_a^d \le 20$ s (colored curves). The hard to soft behavior presented in Fig. \ref{fig3} is confirmed. Moreover, the theoretically predicted time-integrated photon number spectrum $N(E)$ corresponding to the first $20$ s of the ``prompt emission'' (black bold curve) is compared with the data observed by INTEGRAL \citep[green points, see][]{saz,saz2}. This curve is obtained as a convolution of 108 instantaneous spectra, which are enough to get a good agreement with the observed data.}
\label{fig4}
\end{figure}

The simple power-law shape of the integrated spectrum is more evident if we sum tens of instantaneous spectra, as in Fig. \ref{fig4}. In this case we divided the prompt emission in three different time interval, and for each one we integrated on time the energy distribution. The resulting three time-integrated spectra have a clear non-thermal behavior, and still present the characteristic hard to soft transition.

Finally, we integrated the photon number spectrum $N(E)$ over the whole duration of the prompt event (see again Fig. \ref{fig4}): in this way we obtain a typical non-thermal power-law spectrum which results to be in good agreement with the INTEGRAL data \citep[see][]{saz,saz2} and gives a clear evidence of the possibility that the observed GRBs spectra are originated from a thermal emission.

\section{Conclusions}

We show how, applying our model to the GRB 031203, we are able to predict the whole dynamic of the process which originates the GRB emission fixing univocally the two free parameters of the model, $E_{dya}$ and $B$. Moreover, it is possible to obtain the exact temporal structure of the prompt emission taking into account the effective ISM filamentary structure.

The important point we like to emphasize is that we can get both the luminosity emitted in a fixed energy band and the photon number spectrum starting from the hypothesis that the radiation emitted in the GRB process is thermal in the co-moving frame of the expanding pulse. It has been clearly shown that, after the correct space-time transformations, both the time-resolved and the time-integrated spectra in the observer frame strongly differ from a Planckian distribution and have a power-law shape, although they originate from strongly time-varying thermal spectra in the co-moving frame. We obtain a good agreement of our prediction with the photon number spectrum observed by INTEGRAL and, in addition, we predict a specific hard-to-soft behavior in the instantaneous spectra. Due to the possibility of reaching a precise identification of the emission process in GRB afterglows by the observations of the instantaneous spectra, it is hoped that further missions with larger collecting area and higher time resolving power be conceived and a systematic attention be given to closer-by GRB sources.

Despite this GRB is often considered as ``unusual'' \citep{XMM,sod}, in our treatment we are able to explain its low gamma-ray luminosity in a natural way, giving a complete interpretation of all its spectral features. In agreement to what has been concluded by \citet{saz}, it appears to us as a under-energetic GRB ($E_{dya}\approx 10^{50}$ erg), well within the range of applicability of our theory, between $10^{48}$ erg for GRB 980425 \citep{cospar02} and $10^{54}$ erg for GRB 991216 \citep{rubr}.

The precise knowledge we have here acquired on GRB 031203 will help in clarifying the overall astrophysical system GRB 031203 - SN 2003lw - the $2-10$ keV XMM and Chandra data \citep[see e.g.][]{rubr2}.

\acknowledgements

We thank an anonymous referee for important remarks. We thank also S.Y. Sazonov, A.A. Lutovinov and R.A. Sunyaev for their comments on the observational data as well as L. Titarchuk for discussions on the analysis of the convolutions of instantaneous spectra.

\end{document}